# Sound-to-Imagination: An Exploratory Study on Unsupervised Crossmodal Translation Using Diverse Audiovisual Data

**Leonardo A. Fanzeres · Climent Nadeu**

**Abstract** The motivation of our research is to explore the possibilities of automatic sound-to-image (S2I) translation for enabling a human receiver to visually infer the occurrence of sound related events. We expect the computer to 'imagine' the scene from the captured sound, generating original images that picture the sound emitting source. Previous studies on similar topics opted for simplified approaches using data with low content diversity and/or sound class supervision. Differently, we propose to perform unsupervised S2I translation using thousands of distinct and unknown scenes, with slightly pre-cleaned data, just enough to guarantee aural-visual semantic coherence. To that end, we employ conditional generative adversarial networks (GANs) with a deep densely connected generator. Additionally, we present a solution using informativity classifiers to perform quantitative evaluation of the generated images. This enabled us to analyze the influence of network bottleneck variation over the translation, observing a potential trade-off between informativity and pixel space convergence. Despite the complexity of the specified S2I translation task, we were able to generalize the model enough to obtain more than 14%, in average, of interpretable and semantically coherent images translated from unknown sounds.

**Keywords** Computational imagination, crossmodal learning, deep audiovisual learning, generative adversarial networks (GANs), information bottleneck, sound-to-image translation, unsupervised learning.

## 1. Introduction

In the last decades, acoustic event detection (AED) have evolved from borrowing techniques developed for automatic speech recognition (Ballas & Howard, 1987) to the use of deep learning (DL) models (Gencoglu et al., 2014). However, current AED systems are still mostly based on classification processes and, from the perspective of the human receiver, the model output has not changed much. In such systems, the inference result is limited to discrete labels to represent concepts of sounds. A class-based output may be suitable for feeding an automatic audio monitoring system in a restricted acoustic context. Although, considering the complexity of environmental sounds, which constitute a vast and diverse set of concepts, such output can easily mislead the human receiver, inducing to a poor interpretation of the actual sound scene. In a previous study (Fanzeres et al., 2018), deaf participants tested a mobile sound recognition system and expressed their preference for images rather than text to represent sounds in the application. In the present work, as an alternative to sound classification, we explore the possibilities of automatic sound-to-image (S2I) translation to visually inform the occurrence of acoustic events. We propose a system that, given an input audio, is able to 'imagine' the scene with the sound emitting source, generating an original image based on the knowledge acquired through audiovisual learning. Additionally, the system is expected to generate images that are interpretable and semantically coherent with the corresponding acoustic event of the captured audio. Along the text we will eventually refer to such images as informative. This term is used here with a broad meaning, but aligned with the notion of 'informativity' in the context of text translation as described by Neubert and Shreve (1992): "Informativity in the translation process is a measure of the information a translation provides to an L2 reader about L1 events, states, processes, objects, individuals, places and institutions. The original information source was an L1 text intended for L1 audience. Translation opens an information channel between senders and receivers who could not normally inform one another about their respective states of affairs." L1 and L2 mean, respectively, source language and target language.

From pioneer works employing data mining techniques for matching words and image parts (Barnard et al., 2003),

The present work was supported in part by the Brazilian National Council for Scientific and Technological Development (CNPq) under the PhD grant 200884/2015-8. Also, the work was partly supported by the Spanish State Research Agency (AEI) project PID2019-107579RB-I00/AEI/10.13039/501100011033.

Leonardo A. Fanzeres
leonardo.areias@upc.edu

Climent Nadeu
climent.nadeu@upc.edu

Signal Theory and Communications Department, Polytechnic University of Catalonia (UPC), Spain

to recent approaches based on DL models, crossmodal initiatives have gained perspective of new horizons. These proposals share the same essential strategy: create a bridge to connect different modalities. DL models enable an efficient data processing, since they can achieve higher levels of abstraction with automatically learned features. Besides, in the case of convolutional neural networks (CNNs) applied for computer vision, interpretable features may be generated in its inner layers. According to Zhou et al. (2015), semantic parts and objects emerge spontaneously on CNNs trained for scene classification. Gonzalez-Garcia et al. (2018) verified that approximately 10 to 20% of inner features may be composed of interpretable concepts related to textures, materials, semantic parts and objects. Furthermore, Liang et al. (2015) have verified that images and sounds have complementary information about the occurrence of common events on a video stream. Those findings reveal a potential strategy to build the crossmodal bridge, which is based on the assumption that both visual and aural modalities share extractable semantics about acoustic events. For instance, a video of a beach scene may contain images and sounds of waves crashing on the shore. If we can capture the aural-visual correspondence with a tractable and meaningful representation, then we will be able to trace the path towards the aimed direction. Due to the complexity involved, previous studies adopted simplified approaches using data with low content diversity and/or sound class supervision (Chen et al., 2017) (Oh et al., 2019). Differently, we propose to perform crossmodal S2I translation addressing diverse audiovisual content. Although the addressed sonic universe is restricted to five sound classes: Baby cry, Dog, Rail transport, Fireworks and Water flowing, our models were trained with more than eight thousand distinct scenes from a dataset with high inter and intraclass diversity. We also highlight that sound class information was used only for domain restriction and model evaluation, and that our system does not require supervision for training. Besides, our proposal is nearly a real world translation, as we use slightly pre-cleaned data, just enough to guarantee that the acoustic related element/event is present in both aural and visual modalities.

The system we present is an end-to-end solution obtained after the training of an autoencoder to define the audio latent space and a generative adversarial network (GAN) (Goodfellow et al., 2014) with a deep densely connected generator to perform crossmodal translation and synthesize the images. Additionally, we present a solution using informativity classifiers as a way to perform quantitative evaluation of S2I translation. This enabled us to analyze the influence of network bottleneck variation over the translation. In a subtle way, the results indicate a trade-off between informativity and pixel space convergence, obtained respectively from higher and lower audio embedding space dimensionality. Though the specified S2I translation problem is quite challenging, we were able to obtain models that translate more than 14%, in average, of unknown sounds to informative images. As far as we know, this is the first study to tackle S2I translation with such diversity of audiovisual content. Furthermore, we present the techniques that we have developed to address issues like latent space continuity, model generalization, and GANs training stabilization.

This text is organized as follows. In the next section we present a revision of previous works on crossmodal processes. Section 3 explains the difficulties of performing S2I translation. In Section 4, the S2I translator is described. Section 5 presents the results we have obtained with our translator. We then conclude the study presenting our final considerations and suggestions for future works.

## 2. Related Work

The present study proposes a S2I translation system employing DL models (Deng & Yu, 2014) to produce perceptually meaningful and semantically coherent output for the human receiver. Following is a summarized literature review of studies using DL methods for aural-visual crossmodal processes.

Chen et al. (2017) conducted a study on S2I translation using conditional GANs (Mirza & Osindero, 2014). Adopting a translation structure similar to the used by Isola et al. (2017) and Zhu et al. (2017), but addressing a different problem, the authors performed crossmodal content generation in both directions, image-to-sound (I2S) and S2I. The idea is to translate audio tracks of musical solo performances to images of a person playing the corresponding instrument, and vice versa. In a similar way to previously mentioned crossmodal approaches, and partly adapted from the DCGAN architecture (Radford et al., 2016), their system consists of an encoder, a generator and a discriminator. The generator is conditioned with a compressed embedding and regularized with noise injection. The discriminator takes the generated output and pairs it with the compressed embedding from the input modality. Then it computes a score to provide feedback to the net on whether it is a genuine pair of sound and image. Considering S2I translation, their model generated good results using the URMP audiovisual dataset (Li et al., 2019), which contains studio-quality video tracks of uniformly framed people playing instruments with a blue background. However, when employing a more diverse dataset, the quality of the synthesized images dropped considerably. Hao et al. (2018) presented a framework called CMCGAN to handle crossmodal aural-visual mutual generation. Apart from being able to perform I2S and S2I translation also using the URMP dataset, their framework improved quality on crossmodal reverse translation from synthetic image/sound compared to the same task using ground truth image/sound pairs. The authors claim that this advance is due to a better handling of the asymmetry of dimension and structure across different modalities, which is obtained through noise injection. Similarly to the previous work by Chen et al. (2017), the weak point of their approach is the low diversity and the uniformity of the audiovisual content. Duan et al. (2021) employed the same URMP dataset to perform cascade coarse-to-fine S2I translation. Instead of feature embedding, they adopt a supervised approach to keep the crossmodal translation consistent with the high level semantics. Using attention mechanism, class-based loss in all learning stages and a residual class label to guide the finer image generation, they were able to improve significantly the results obtained by Chen et al. (2017) and Hao et al. (2018). Differently from these approaches, our work proposes to address the S2I translation problem without supervision and employing larger and more diverse audiovisual datasets.

Wan et al. (2019) present results of S2I translation using conditional GANs trained on video data. Unlike ours, their approach is completely supervised. Apart from extracting the audio feature vector from SoundNet (Aytar et al., 2016), the generator and the discriminator are both trained with an auxiliary classifier to improve the semantic coherence between the generated image and the respective input sound. Also, they propose a sound-image similarity score to improve discriminator training. Similarly to the work of Chen et al. (2017), their S2I translator generates relatively informative images addressing audiovisual data that present low content diversity and background uniformity. The classes used are: Baseball, Dam, Plane, Soccer, and Speedboat. Using a subset of the same data, Yang et al. (2020) present a S2I translator based on a stacked GAN architecture. Like the work of Wan et al. (2019), their approach is entirely supervised. They also use SoundNet extracted audio features for the generator input and an auxiliary classifier for GANs training. Besides, they reverse the translation with an I2S net to verify the audio content consistency between the original audio embedding and the reversed one. The authors present the results of S2I translation from two sound classes, Baseball and Soccer, obtaining informative images. However, the dataset used for the experiment consists of only 2065 sound-image training pairs. Moreover, in these two studies, the weak point of the experiments is that the split of the data does not ensure that sound-image pairs of the same video will not appear in both training and test sets. This procedure compromises the evaluation of the system, since the model may be tested with sounds from known scenes.

Another S2I task that has already been explored is the generation of images of faces from a given speech. Duarte et al. (2019) present an end-to-end solution called Wav2Pix. They address the crossmodal problem using a GAN conditioned on an audio embedding extracted from speech. The model is able to generate realistic and diverse face images. However, their approach requires a clean dataset with precisely framed faces and high audiovisual quality. Besides, to obtain acceptable results, their generator needs to be conditioned on known voices. Oh et al. (2019) present a model called Speech2Face to address a similar problem. They train an encoder to match the visual embeddings to those generated by a pre-trained face recognition network (Parkhi et al., 2015). The decoding process is performed using a separately trained reconstruction model (Cole et al., 2017) which generates images of faces in a canonical form, i.e. precisely framed, frontal-positioned and with neutral expression. Unlike our approach, these works address a specific domain, which significantly reduces audiovisual content diversity.

Chatterjee and Cherian (2020) present a framework called Sound2Sight to perform video frames generation conditioned on the audio track and past frames. Briefly described, their framework follows an encoder-decoder auto-regressive generator architecture, generating one video frame at a time using two long short-term memory (LSTM) networks. Differently from our approach, Sound2Sight does not process pure S2I translation, since the visual modality is also present in the input. This constraint helps to generate more plausible images, but makes the entire process be closer to a multimodal than to a crossmodal task. Moreover, the conducted experiment employs three datasets separately, each one consisting of specific audiovisual content, therefore presenting low diversity.

Also addressing a S2I task, Shim et al. (2021) propose an end-to-end solution for generating images of birds conditioned on call sounds of correspondent species. After training a sound classifier, they obtain the audio embedding which is then input to a conditional GAN. Different from our proposal, they use a class based encoder. Besides, the adversarial training is also supervised, using a discriminator that is trained to output both generated images realness and species label. Moreover, their study is domain specific, addressing limited audiovisual diversity. In another study on bird sounds, Hao et al. (2021) present AECMCGAN, a framework based on their previous work (Hao et al., 2018), to perform both I2S and S2I translation with the addition of attention modules to model intra and inter-modality global dependencies. They obtained improved results compared to former studies using their own dataset for birds sound-image crossmodal translation and a subset of the previously mentioned URMP dataset. However, both datasets are domain specific, which limits content diversity. Another downside of these two studies on birds crossmodal translation is that the split of the data do not ensure that the model will not be tested with sounds from known audio streams. This strategy, as mentioned before, compromises the evaluation of the translation.

Furthermore, a recent work by Zhu et al. (2021) that is worth reading, presents a survey on deep audiovisual learning. The authors organize the text into four main topics: audiovisual separation and localization, audiovisual corresponding learning, audio and visual generation, and audiovisual representation. The study covers some works mentioned above among others.

## 3. Inherent Challenges in S2I Translation Processes

S2I translation has in common with other crossmodal tasks the problem of finding the semantic correlation between the two modalities. In our case, the system has to link the acoustic events present in the audio stream to the semantically correlated elements of interest in the visual modality. While it is an easy and intuitive process for humans to learn that semantic correlation between images and sounds, it becomes a challenging task for machines due in large part to the discrepancy between the audio waveform domain and the image RGB color domain (H. Zhu et al., 2021). For instance, there is a common conceptual or semantic entity between a foreground dog bark sound in a background acoustic environment and a picture that includes a dog in a salient position, but the heterogeneity of the representation of the dog entity in those two domains is an obstacle to perform S2I translation.

Due to its characteristics, the crossmodal generation involved in S2I translation turns out to be a task that belongs to a broader field of artificial intelligence (AI), that is computational imagination (Mahadevan, 2018) (Davies, 2020). And its complexity is also a consequence of the fact that such artificial systems aim to simulate an ability that can be considered exclusive of the human being. Based on Stevenson's work (Stevenson, 2003), Beaney (2010) discusses an alignment between philosophers regarding a possible definition of 'imagining', which can be conceived as 'thinking of something that is not present to the senses'. In the context of our study, the missing part is the entire visual modality. When we ask the translator to generate an image based on the input sound, e.g. a baby crying, we intend the output to be a complete and particular scene with a baby crying, catching all possible information from the audio signal to picture an informative image, not a prototype of a baby standardly positioned in a neutral background. And since the input sound will most likely be different from any sound known by the translator, we assume that the generated image will probably not look like the visual surrounding corresponding to the captured sound. Besides, many elements of the original scene may leave no trace on the audio signal, as, for instance, the color of the baby's clothes. Thus, as in a sound-to-imagination mental process, we expect the computer to use its audiovisual knowledge to 'imagine' an approximate scene with the sound emitting source, as well as related elements that can contribute to picture an informative image.

Another inherent characteristic of crossmodal generation processes is computational creativity, which is fundamental to provide original outputs. To 'imagine' the surrounding scene corresponding to the sound, the system may need to blend known images, gathering visual 'memories' from the imagery acquired during the training phase, so as to create an original image. To reach this goal, the system must analyze the input sound, involving the encoded patterns that have been learned from the training sounds, to generate a representation of it that effectively drives the 'creation' of the output image. The outlined process follows ideas frequently exposed in philosophy, psychology and cognitive science about the interrelation between imagination, perception, memory, knowledge, and creativity (Beaney, 2010). Pereira and Cardoso (2002) emphasize the importance of extending the established AI techniques to improve divergence/convergence abilities (Guilford, 1967) (Gabora, 2019) of algorithms in order to achieve 'creativity'. Among other AI approaches, genetic algorithms are probably the first to employ convergence/divergence methods to obtain original results, but mostly limited to explore narrower knowledge spaces (Pereira & Cardoso, 2002). More recent AI techniques such as GANs are capable of exploring wider spaces. The adaptation of this architecture with a conditioned generator (Mirza & Osindero, 2014) is a common approach in current studies on crossmodal tasks. However, despite its effectiveness on producing realistic results, GANs are known to be unstable due to adversarial training (Mescheder et al., 2017) (Mescheder et al., 2018). A solution presented by Radford et al. (2016) named Deep Convolutional GANs (DCGAN) helps to overcome this behavior. Their approach consists of a set of architectural constraints that has shown to stabilize GANs output in most settings. Although other forms of instability like filters collapse still remain (Radford et al., 2016). Those issues take us to a fundamental problem on training generative models, that is the need to increase the synthesized content diversity (J.-Y. Zhu, Zhang, et al., 2017) (Mao et al., 2019). Solving this is especially important when the quest for creativity is involved, as in our case. A greater diversity of training data can indeed improve the model's ability to generate diverse output (Tobin et al., 2017). On the other hand, excessive diversity may hamper the generaliza-tion of the model. We are indeed aware of the difficulty to tackle data diversity, both in training and in data generation. In this work, we employ diverse audiovisual content for training the translator, and we need to obtain a model with enough generalizability to 'imagine' informative images from unknown sounds. Also, the model must be sufficiently 'creative' to output diverse and original images. It is a really difficult task, and, despite the low percentage of informative images obtained (around 14%), we show that it is possible to take steps towards the S2I translation objective.

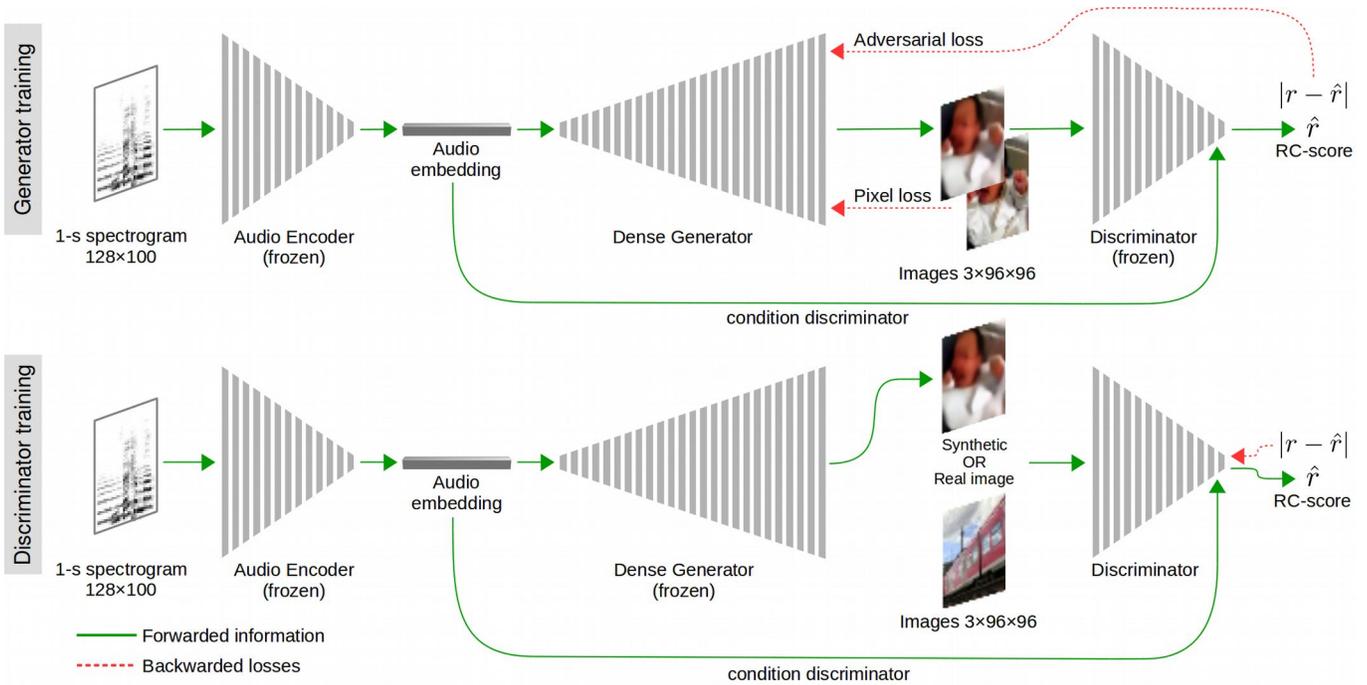

**Fig. 1** S2I translator training scheme

We train our translator with 48,945 sound-image pairs, extracted from more than eight thousand different scenes with diverse audiovisual content, which are representative of the kind of visual output we expect the translator to generate. Besides, we test the translator exclusively with scenes unknown by the model. These criteria make our study to distinguish from previous works which adopted more simplified approaches, without addressing content diversity and/or using sound class supervision, as explained in the previous section. In some of the studies mentioned, the task being performed is similar to an image retrieval process, where the system fetches in a database the image that best fits the query. The exploration we have done, with the corresponding solutions, and the S2I translator we present are a modest attempt to address such big challenge. Potential alternative approaches include attention-based methods (Vaswani et al., 2017) (Zhang et al., 2019), causal reasoning (Pearl, 2009) (Walker & Gopnik, 2013), prior knowledge made explicit by rules and constraints (Stewart & Ermon, 2017), bags of acoustic events (Grzeszick et al., 2017) and/or visual elements (Kato & Harada, 2014), sound separation (Kavalerov et al., 2019) and/or image segmentation (Long et al., 2015), and also different levels of supervision. Those techniques impose restrictions that could even make the translator produce more realistic results. However, this limiting structure of the data may lead the system to get in conflict with the objectives of the translation we propose, which demands a commitment to diversity and imagination. Instead of explicit human-generated knowledge, we explore the power of deep neural networks (DNN), and GANs in particular, to model the aimed S2I translator. Due to the complexity of the task, it has been a tricky process to design and train these DNN models. Without minimizing the problem, except for constraining the number of sound classes used, we adopt a clear exploratory approach, presenting a solution that gives clues on how to face the challenge. Nonetheless, we also provide brief descriptions and suggest possible explanations for the observed phenomena. Furthermore, we chose to perform this task without sound class supervision in order to make the scaling of the model more feasible in the future. This strategy also avoids class biased results, allowing the translator to freely use the acquired multimodal knowledge, without any eventual restriction that could be imposed by a supervised approach.

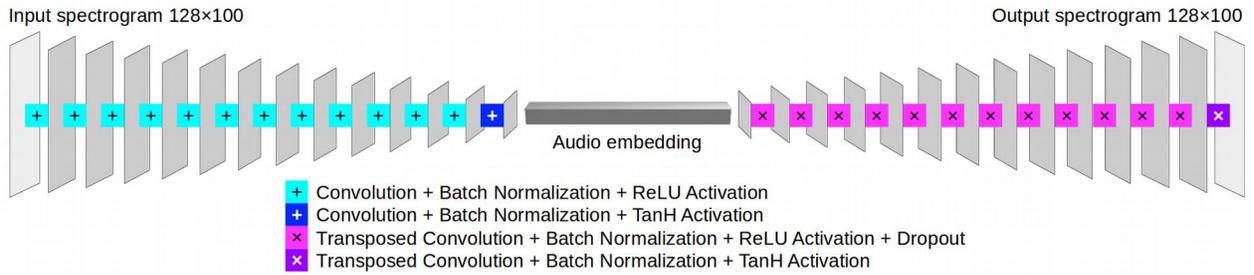

**Fig. 2** Audio autoencoder architecture

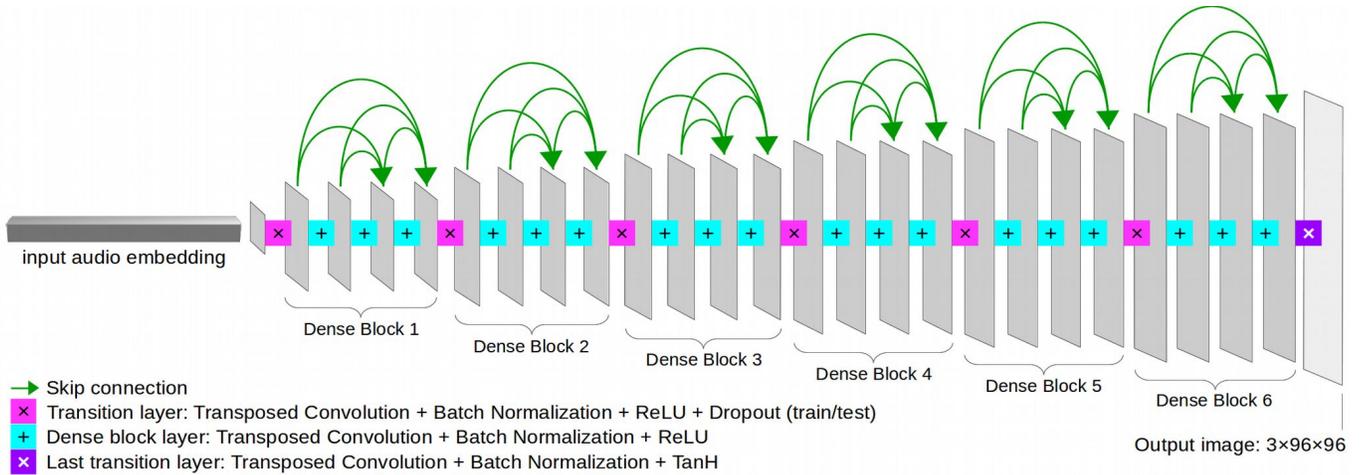

**Fig. 3** Generator architecture inspired in DenseNet

## 4. Sound-to-Image Translator

We developed an end-to-end solution that includes a convolutional audio encoder and a conditioned deep densely connected crossmodal generator trained with an also conditioned discriminator. As exposed in the previous section, S2I translation is a complex task and, since the beginning of this research, we were aware that dealing with content diversity would be a major challenge. This fact guided many decisions regarding topics like neural networks architecture, model regularization and training algorithm, which are detailed in this section.

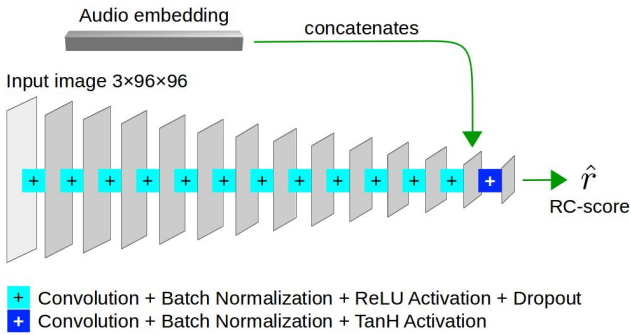

**Fig. 4** Discriminator architecture

### 4.1. Overview

The S2I translator training is outlined in Figure 1. Firstly, an audio autoencoder is trained with log-Mel spectrograms computed from 1-s audio segments. Then, we use the frozen encoder to extract the audio embeddings that will be forwarded to the generator. During the training phase, the generator will try to fool the discriminator, which is trained once for every five updates of the generator. At this point, the discriminator will receive balanced batches of real and synthetic images and their corresponding target scores to model visual feature extraction. An aural-visual coherence checking is done through concatenating the source audio embedding to the input of the discriminator's last layer, fusing aural and visual modalities. This will enable the discriminator to jointly check both the realness of the generated images and their semantic coherence with the corresponding audio, and output what we call a realness-and-coherence score (RC-score).

### 4.2. Network Architecture

As it can be seen in Figure 2, the audio autoencoder consists of a mirrored architecture of 26 convolutional layers, each one followed by batch normalization (BN). Both encoder and decoder inner layers are activated by rectified linear units (ReLU) while the last layer employs a hyperbolic tangent function (TanH). The decoder part also includes a dropout regularization on each inner layer to prevent overfitting.

With respect to the generator architecture, the first noteworthy improvement from the baseline S2I translator was due to the use of a deeper 25-layer architecture, specially when skip connections have been applied to make a dense generator. The final architecture is inspired in DenseNet (Huang et al., 2017) (Fig. 3). Compared to the initial 13-layer sequential generator that had a structure similar to the audio decoder, the new deeper and dense architecture resulted in a significant increase in the quality of the images generated. The convolutional layers of the generator are followed by BN and ReLU activation, except for the output layer which uses TanH. In addition, we include a dropout regularization between each dense block to improve the generalization of the model, as well as to avoid deterministic inference through its application also at test time, as done by Gal and Ghahramani (2016) and Isola et al. (2017).

Regarding the discriminator architecture (Fig. 4), except for the input and the feature maps shape, it shares its structure with the audio encoder, adding the audio embedding conditioning on the input of the last convolutional layer and applying also dropout regularization after each inner layer. The last convolutional layer directly outputs a scalar corresponding to the RC-score.

### 4.3. Method

Considering only input and output variables, the data of the networks are detailed as follows. We define a set of data triples sound–embedding–image $\{S_i, x_i, Y_i\}$, consisting of spectrograms $S_i \in \mathbb{R}^{h \times w}$, audio feature vectors $x_i \in \mathbb{R}^f$ and real color images $Y_i \in \mathbb{R}^{h \times w \times c}$, where each element of the triple corresponds to the same $i^{th}$ acoustic event. All the real numbers are limited to the interval $[-1, 1]$, since pixel values are normalized to the mentioned range before entering the network, and audio embeddings, as well as the RC-score $r \in \mathbb{R}$, both activated by TanH, fit the same interval. Regarding dimensions: $h$ and $w$ are height and width of spectrograms or images, $c$ is the number of channels of color images, and $f$ is the dimensionality of the audio feature space. The audio encoder $A_E$ and the audio decoder $A_D$ are defined, respectively, as the transformations $A_E(S) : \mathbb{R}^{h \times w} \to \mathbb{R}^f$ and $A_D(x) : \mathbb{R}^f \to \mathbb{R}^{h \times w}$. From the side of the adversarial networks, the generator $G$ and the discriminator $D$ are defined, respectively, as $G(x) : \mathbb{R}^f \to \mathbb{R}^{h \times w \times c}$ and $D(Y \vee G(x), x) : \mathbb{R}^{h \times w \times c + f} \to \mathbb{R}$, where $Y$ denotes the real image that is entered to the discriminator alternating with the synthetic image from $G(x)$.

With respect to the computation of loss functions, we opted for using the mean-squared error (MSE) on the pixel space for measuring the distance between the target and generated spectrogram/image during the training of the audio autoencoder and the generator. This strategy resulted in an unbiased and scalable solution that allowed us to do extensive testing and empirically learn from the behavior of the translator, thus providing essential experience to improve the architecture and tune the networks.

The optimization of the autoencoder networks $A_E$ and $A_D$ is performed minimizing the pixel loss $L_A(S, \hat{S})$ defined in Equation 1, where $b$ is the batch size, and, as stated earlier, $h$ and $w$ are height and width of spectrograms. The loss is computed as the MSE between the target spectrogram $S$ and the generated one $\hat{S} \leftarrow A_D(A_E(S))$. We include the batch iteration in the equations since this is how the losses are effectively computed, being averaged among all instances of the batch.

$$L_A(S, \hat{S}) = \frac{1}{bhw} \sum_{i=1}^{b} \sum_{j=1}^{h} \sum_{k=1}^{w} (S_{ijk} - \hat{S}_{ijk})^2 \quad (1)$$

The discriminator $D$ is optimized through the minimization of the score loss $L_D(r, \hat{r})$ defined in Equation 2, which is calculated from the batch-averaged MSE between the output RC-score $\hat{r} \leftarrow D(Y \vee G(x), x)$ and the target RC-score $r$, which can be the maximum or the minimum score, depending on whether the input image is real or synthetic, respectively.

$$L_D(r, \hat{r}) = \frac{1}{b} \sum_{i=1}^{b} (r - \hat{r}_i)^2 \quad (2)$$

The optimization of the generator $G$ is shaped by two objectives. The first aims to minimize the pixel loss $L_G(Y, \hat{Y})$ defined in Equation 3, where $c$, as stated earlier, is the number of channels of color images. The loss is computed as the MSE between the target image $Y$ and the generated one $\hat{Y} \leftarrow G(x)$.

$$L_G(Y, \hat{Y}) = \frac{1}{bchw} \sum_{i=1}^{b} \sum_{j=1}^{c} \sum_{k=1}^{h} \sum_{l=1}^{w} (Y_{ijkl} - \hat{Y}_{ijkl})^2 \quad (3)$$

The second objective aims to minimize the adversarial loss based on the RC-score obtained from the discriminator $D$. We implemented a moving-average adversarial loss $L_G^{ma}(r_{max}, \hat{r})$ defined in Equations 4 and 5, where $r_{max}$ is the maximum RC-score value, $t$ is the current epoch number, $\overline{L}_{Gi}$ is the average adversarial loss for epoch $i$ and $k$ is the number of averaged data-points. When used to train the generator instead of the current batch loss $L_G(r_{max}, \hat{r})$ (Eq. 4), the moving-average loss significantly attenuated the adversarial training instability. In our case, this is especially important because the ratio of generator/discriminator training update, which is 5, is higher than usual and provoked even more instability during GANs training.

8                                             Leonardo A. Fanzeres & Climent Nadeu

$$L_G(r_{max}, \hat{r}) = \frac{1}{b}\sum_{i=1}^{b}(r_{max} - \hat{r}_i)^2 \quad (4)$$

$$L_G^{ma}(r_{max}, \hat{r}) = \frac{L_G(r_{max}, \hat{r}) + \sum_{i=t-k+1}^{t-1}\overline{L}_{Gi}}{k} \quad (5)$$

The final generator loss is expressed in Equation 6. The adversarial loss is scaled by a factor $\lambda$ to balance the mean amplitude of the two terms.

$$L_G = L_G(Y, \hat{Y}) + \lambda L_G^{ma}(r_{max}, \hat{r}) \quad (6)$$

Leaving apart the audio encoder training, we present below an algorithm describing the main steps of the training of our translator:

---

**Require:** $b$, the batch size; $n_g$, the number of training iterations of the generator; $n_{gd}$, the number of training iterations of the generator per discriminator training; $r_{min}$, the minimum RC-score value; $r_{max}$, the maximum RC-score value; $\lambda$, the adversarial loss scale factor.

1: **for** $n_g$ iterations **do**
2:     Get $b$ spectrograms from stored data:
        $S_{batch}\{S_1, S_2, ..., S_b\} \leftarrow data$
3:     **with** $A_E$ frozen
4:         Get $b$ audio embeddings from the audio encoder:
            $x_{batch}\{x_1, x_2, \ldots, x_b\} \leftarrow A_E(S_{batch})$
5:     **end with**
6:     Get $b$ real images from stored data:
        $Y_{batch}\{Y_1, Y_2, ..., Y_b\} \leftarrow data$
7:     **if** current iteration number is multiple of $n_{gd}$ **then**
8:         **with** $G$ frozen
9:             Update the discriminator to minimize:
            $(L_D(r_{max}, D(Y_{batch}, x_{batch}))$
            $+L_D(r_{min}, D(G(x_{batch}), x_{batch})))$
10:        **end with**
11:    **end if**
12:    **with** $D$ frozen
13:        Update the generator to minimize:
           $(L_G(Y_{batch}, G(x_{batch}))$
           $+\lambda L_G^{ma}(r_{max}, D(G(x_{batch}), x_{batch})))$
14:    **end with**
15: **end for**

---

## 5. Experiments

In this section we explain the heuristics of our approach, detailing training strategies and the datasets used for the experiments. Also, we present our solution for evaluating the translated images using informativity classifiers. We complete the section providing both quantitative and qualitative evaluation of the S2I translation results. Further information about the experiments and the code implemented for the networks training are available at https://purl.org/s2i.

### 5.1. Data Used

The AudioSet[1], described by Gemmeke et al. (2017), consists of a large-scale audiovisual dataset of manually annotated acoustic events. Starting from the literature and manual curation, the authors defined a structured hierarchical ontology of 632 audio classes to collect data from human labelers. The goal of their task was to probe the presence of specific audio classes in 10-s segments of YouTube videos. The complete dataset contains more than 2 million videos and the labeled segments employ part of the AudioSet ontology. The provided data is characterized by highly diverse audiovisual content. Dealing with such variety makes our study to distinguish from previous works. Despite that we use only five sound classes, the sound-image pairs are extracted from more than eight thousand distinct scenes, resulting in high inter and intraclass diversity. For training and testing our S2I translator, we employed an AudioSet subset named VEGAS, which was made available by Zhou et al. (2018), for their study on crossmodal image-to-sound translation. This dataset provides cleaner data, where the start and end of addressed acoustic events are precisely annotated. Besides, the tracks have been inspected to verify if the element/event of interest for the video clip were present in both visual and aural modalities, and non-matching segments were removed. The complete VEGAS dataset consists of 28,109 videos of 10-s maximal duration distributed in 10 sound classes, among which we use five: Baby crying, Dog, Rail transport, Fireworks, and Water flowing. The original

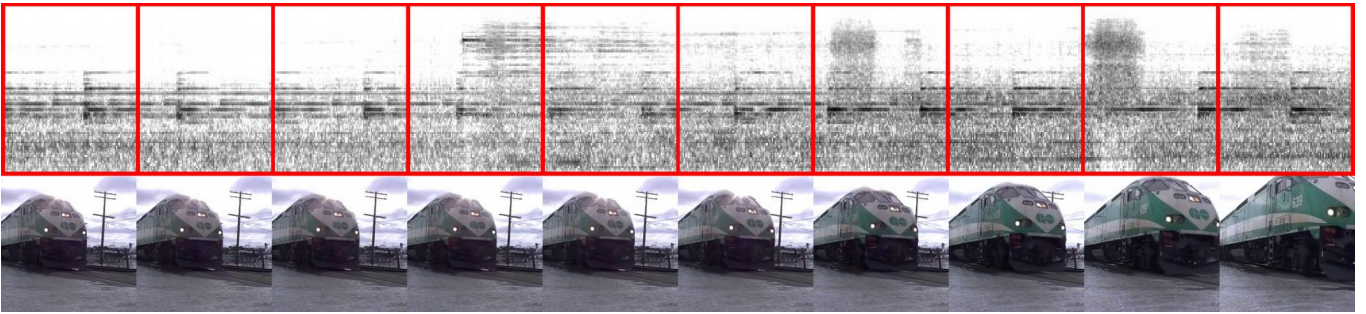

**Fig. 5**   1-s Mel spectrogram segments and respective central frames extracted from a video track of the class Rail transport.

1 https://g.co/audioset

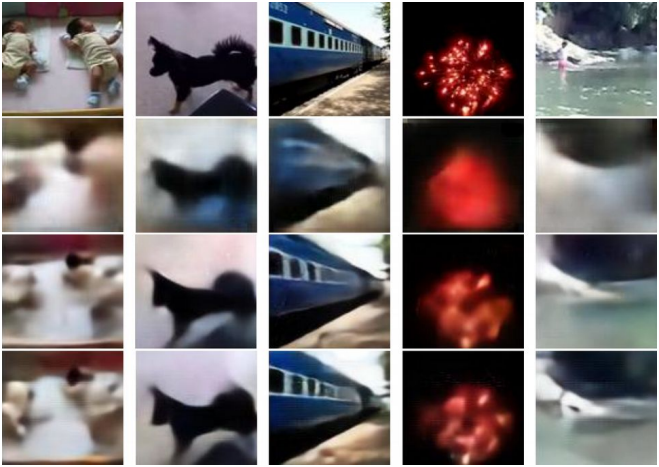

**Fig. 6** Illustrative comparison between ground truth (1st row on top), and images generated by the sequential 13-layer (2nd row), sequential 25-layer (3rd row), and the dense 25-layer generator (4th row) using known sound-image pairs.

VEGAS dataset is unbalanced, therefore, to avoid class biasing, the same number of segments is used for all classes. Table 1 presents the number of original video tracks for each sound class, and their respective 1-s segments for training, validation and test sets.

For audio data, log-Mel spectrograms were computed according to the following procedure: the signal is split into 25-ms frames, with 15-ms overlap; a Hamming window is applied to the frames and the Short-Time Fourier-Transform (STFT) is computed; its squared magnitude is integrated in 128 sub-bands using triangular weights according to a non-linear Mel-scale, and the logarithm of those sub-band energies is computed. For a 1-s audio segment, a matrix of 100×128 is obtained. One of those spectrograms segmentation and the respective frames from the original video are illustrated in Figure 5. For visual data, images are extracted from the central frame of the corresponding 1-s video segment. Before being loaded into the neural network, the extracted images are square cropped in the center and then resized to 96×96 pixels.

**Table 1** Sound classes and their respective number of video tracks and 1-s segments for training, validation and test sets.

| Sound classes | # of original video tracks (max. 10 s) | # of 1-s video segments | | |
|---|---|---|---|---|
| | | Training | Validation | Test |
| Baby cry | 2059 | 9789 | 1115 | 1365 |
| Dog | 2785 | 9789 | 1115 | 1365 |
| Fireworks | 3115 | 9789 | 1115 | 1365 |
| Rail transport | 3259 | 9789 | 1115 | 1365 |
| Water flowing | 2924 | 9789 | 1115 | 1365 |
| Total | 14142 | 48945 | 5575 | 6825 |

### 5.2. Preliminary Exploration and Training Details

As mentioned in Section 4.2, a fundamental improvement in the quality of translated images was due to the evolution of the generator from a 13-layer sequential architecture to a 25-layer densely connected structure. The deeper architecture composed of dense blocks inspired in DenseNet (Huang et al., 2017) enabled the translator to picture shapes, colors and sharp edges much closer to the original ground truth training images. From this point, it was possible to achieve a reasonable quality of translation without overfitting the model. Since it is difficult to make a quality comparison using test data due to visual decoupling between target and translated images, we show in Figure 6, as illustration, images generated using training data, i.e. known sound-image pairs. We present results from the three architectures of the generator: sequential 13-layer, sequential 25-layer and dense 25-layer.

With respect to the audio autoencoder training, we managed to soften the latent space through dropout regularization on the audio decoder, which improved generalization significantly. In the case of the generator, as mentioned in Section 4.2, we included test time dropout regularization between each dense block. This strategy, in addition to helping generalize the model, induces stochasticity on the generator, enabling greater visual variety in translation. In fact, applying this technique resulted in a noticeable enhancement of generated images diversity. However, apparently, test time dropout improved model generalization only to a certain extent, and translated images were often non-informative. Actually, evaluating the generalizability of the generator is not an easy task, since synthesized images most likely do not share the visual structure with the images that correspond to the input sound. This visual decoupling implies that even if generated images are semantically coherent with the original sound, visual elements will not necessarily appear on the same position as in the target images, which means that the pixel loss obtained from the test set will be useless. In fact, the visual matching hardly ever happens, and when it does, it is an approximate match. Aware of those limitations, we decided to focus our analysis on two characteristics that we believe are suitable for evaluating the quality of the generated images: interpretability and semantic coherence, which can be summarized in the term informativity.

Along our exploration, we also observed that the activation functions of both the generator and the discriminator were playing a prominent role in the process. And, as reported by Glorot et al. in image classification and text sentiment analysis tasks (2011a) as well as in domain adaptation (2011b), network activation sparsity improved generalization. The same effect was verified in our translator during networks tunning. Based on that, we employed ReLU activation for all inner layers of both autoencoder and GANs. The network sparsity provoked by ReLU layers helped to regularize the model, allowing to capture the essential semantic information from input sounds and keep it until the lower levels of abstraction of the generator model, close to the effective synthesis of the image at the end of the net. This explains the better generalization observed, with a noticeable increase of the informativity of generated images. Training the translator with Leaky ReLU or exponential linear units (ELU) activation resulted less stable and models output frequently passed from blurry to sharp but rather abstract images. Although producing increased valid details for training data, models activated by these functions were overfitted, generating more non-interpretable images full of visual artifacts when translating unknown sounds. In fact, some models activated by Leaky ReLU were not able to output one single informative image.

Also, we used an equal number of real and synthetic images, not being necessary to variate their ratio as indicated by Lucas et al. (2018) since we did not experience noticeable mode collapse issues, except during the very beginning of GANs tunning. For all networks we employed the Xavier initialization method, which Glorot and Bengio (2010) referred to as 'normalized initialization'. Therefore, weights $W$ for each network layer are sampled from the random uniform distribution defined in Equation 7, where $n_i$ is the number of incoming network connections and $n_{i+1}$ is the number of outgoing network connections for the layer $i$.

$$W \sim U\left(-\frac{\sqrt{6}}{\sqrt{n_i + n_{i+1}}}, \frac{\sqrt{6}}{\sqrt{n_i + n_{i+1}}}\right) \quad (7)$$

Regarding algorithm hyperparameters, all autoencoders of the five different embedding dimensions were trained with an initial learning rate of 0.05 and momentum 0.9. While both generator and discriminator networks from all embedding dimensions employed an initial learning rate of 0.1 and momentum 0.5. The scale factor $\lambda$ of the generator adversarial loss (Eq. 6) was set to 0.1 and the discriminator was trained once for every five updates of the generator. We adopted a mini-batch gradient descent optimization for all training, using a batch size of 64 for each network update. The entire system is implemented in Python version 3.5.2. The machine learning related code was implemented using PyTorch library version 1.1.0. All training processing of the models was run in a Supermicro[2] SYS-7048GR-TR server allocating 160 gigabytes of RAM and two Intel[3] Xeon E5-2670 processors at 2.30GHz. The execution was accelerated using a GeForce[4] Titan Xp GPU accessed with CUDA[5] platform version 9.0.176.

### 5.3. Informativity Classifiers

As an additional contribution, we present a solution using classifiers to infer whether the translated images are interpretable and semantically coherent, or briefly, if they are informative or not.

Evaluating the true quality of the S2I translation turned out to be an additional difficulty during the experiment. Since class information was available, we initially opted to use ordinary image classifiers to evaluate our translator. The idea was to verify if the generated image would be classified as the original input sound class. However this alternative proved to be ineffective since the reported classification scores were unrealistically high. As exposed in Section 3, S2I translation is a challenging task and generated images are mostly non-informative. During the tests, more than 50% of generated images were reported as matching the original sound class. But looking to the images it was clear that this did not represent the real performance of the translator, even using the best generator models. Two facts made the evaluation of the S2I translation more difficult, producing biased results on classification. First, the low percentage of informative images, which is expected due to the inherent difficulty of the process. This turned out to be a problem, as the large amount of non-informative images easily distorted the classification result. For instance, if these non-informative images were randomly classified among the five sound classes, they would considerably inflate the overall effective accuracy. Second, the fact that even informative images were essentially different from real images. They were usually less sharpen and presented lower visual diversity compared to the real images with which the classifiers were trained. We even tried to analyze the output vector of the softmax function from the visual classifier in an attempt to find correlation between the class distribution and the reliability of classification, but no improvement was verified.

After the failed attempts mentioned above, we proposed to classify images as informative and non-informative in order to quantitatively evaluate the S2I translator performance. It was necessary to train one informativity classifier for each sound class, since general classifiers performed poorly. Besides, such general classification would not benefit the experiment in any way, given that our goal was uniquely to report the translators performance as accurately as possible. Thus, we trained five informativity classifiers using informative and non-informative synthetic images translated from validation sounds. It is important to keep in mind that there are two different visual data sectioning here. One that divides the set of images in five sound classes, no matter if they are real or synthetic images, while the other divides it in informative and non-informative images, and is employed only for evaluating the quality of generated images. Considering these data perspectives, we built five balanced datasets totalizing 5000 synthetic images selected among the outputs of 17 previously trained S2I translator models. Each sound class dataset consists of 1000 images (800 for training and 200 for testing) evenly distributed

2, 3, 4, 5 Trademarked

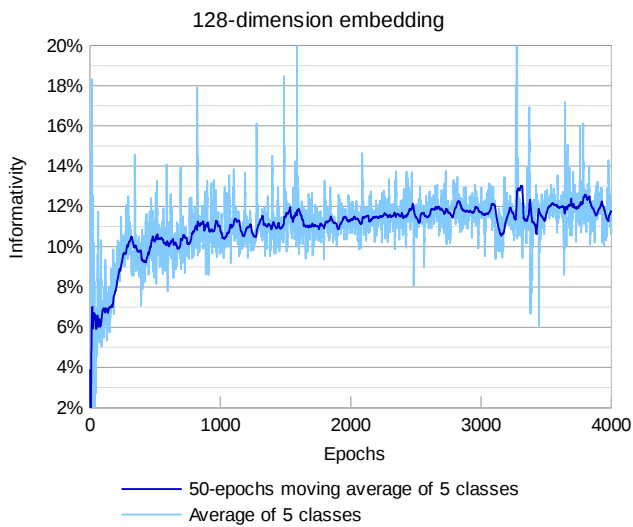

**Fig. 7** Translator's informativity from a 128-dimension embedding.

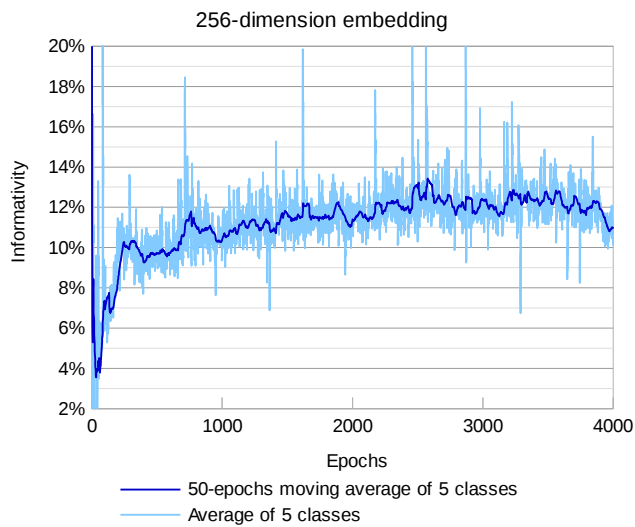

**Fig. 8** Translator's informativity from a 256-dimension embedding.

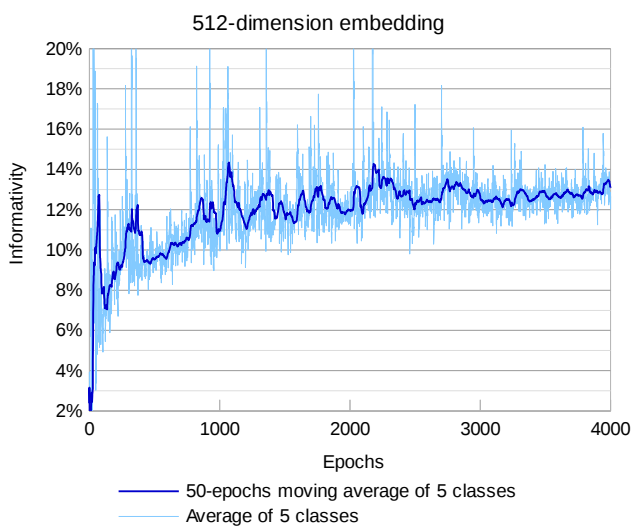

**Fig. 9** Translator's informativity from a 512-dimension embedding.

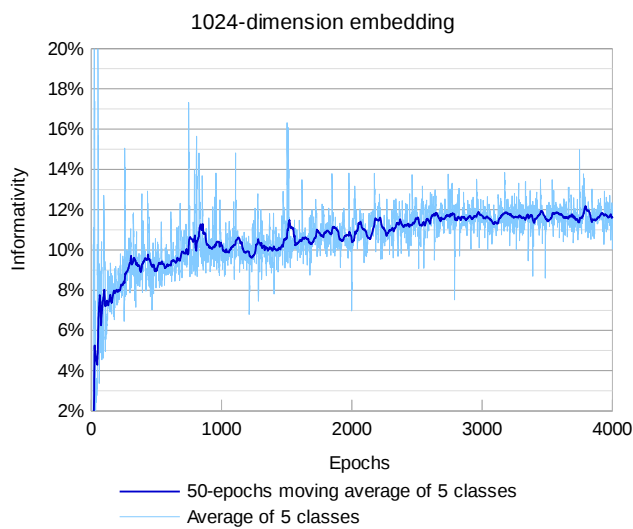

**Fig. 10** Translator's informativity from a 1024-dimension embedding.

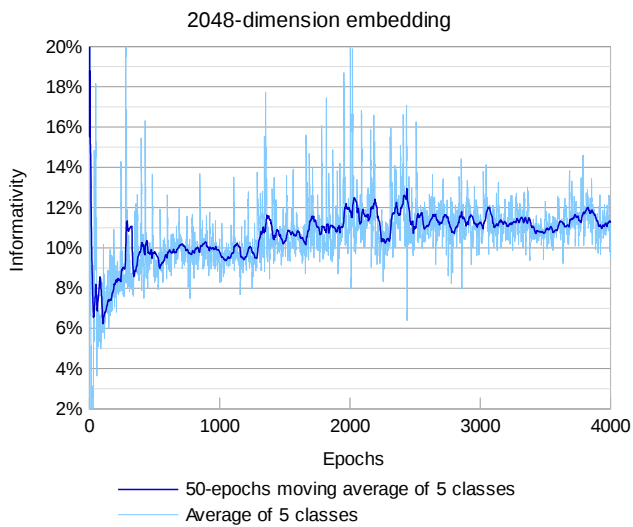

**Fig. 11** Translator's informativity from a 2048-dimension embedding.

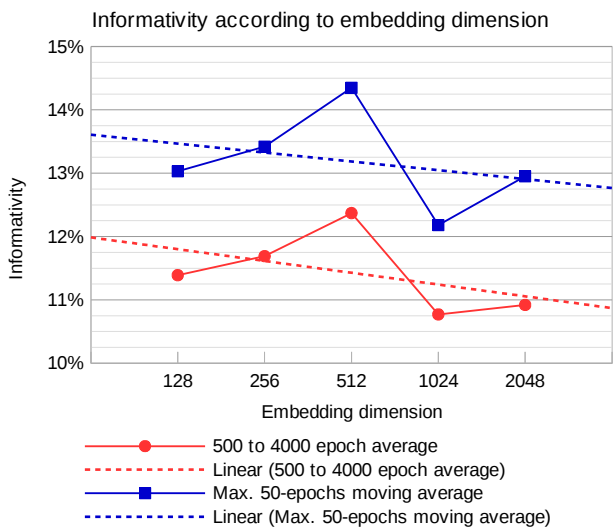

**Fig. 12** Translator's informativity from 5 different embedding dimensions.

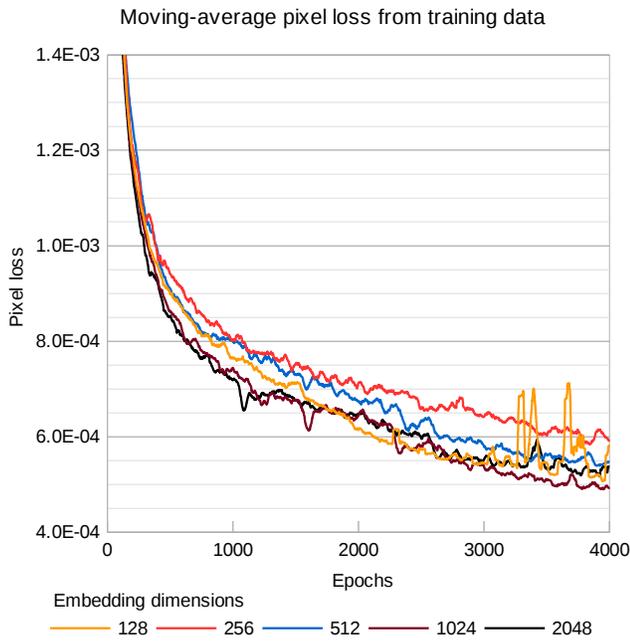

**Fig. 13** 50-epochs moving-average pixel loss from training data.

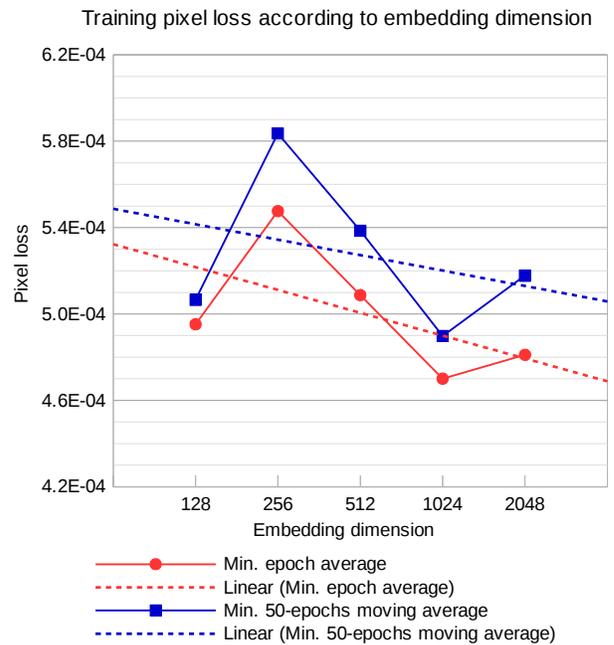

**Fig. 14** Average pixel loss from training data.

between informative and non-informative classes. Since at that moment we still did not have any informativity classifier, the screening of translator models was based on the reported pixel loss and visual class match of generated images, and image selection was based on subjective evaluation.

The S2I translators performed differently for each sound class. For instance, when we employed the translator to generate images from sounds of water flowing we could achieve approximately 18% of informative images, while when using dog sounds, the best performance was about 6%. For this reason we had to use more translator models for classes with lower performances until we completed the set of 500 informative images for each sound class. Regarding the non-informative class, although only one model would be sufficient to complete the 500 images set, we maintained the same number of images per model to prevent biasing the informativity classifier. This is necessary because image generation models may leave a fingerprint on the output. If a different model distribution was used, subtle artifact patterns (Odena et al., 2016) or even the blur level could give clues of the input image's original class.

We highlight that these informativity classifiers were used only to report the performance of the translator, therefore, supervision information is not used for the translator's training. The architecture of the classifiers is composed of a CNN with five convolutional layers for visual features extraction and two fully connected layers for classification. Batch normalization, ReLU activation and dropout (for training) are applied after each layer, except the last fully connected, where the log probabilities are computed from the output of a softmax function. All five classifiers were trained with an initial learning rate of 0.001, momentum 0.9 and weight decay of $5^{-5}$, obtaining models with the following accuracies – Baby cry: 80%; Dog: 80%; Rail transport: 84%; Fireworks: 82%; and Water flowing: 81.5%.

### 5.4. S2I Translation Results

We present here the performance of different S2I translator models regarding variation of audio embedding space dimensionality. In addition, we provide and analyze a selection of images that is representative of the S2I translation achieved quality.

#### 5.4.1. Quantitative Evaluation

Employing the set of five informativity classifiers we were able to make an extensive comparison of the performance of different S2I translators. Using the architecture presented in Section 4.2, the best performance of our translator was obtained with an audio embedding dimension of 512, achieving more than 14% average informativity among the five sound classes. Translators informativity history is reported in Figures 7 to 11. Due to oscillation of the obtained informativity along the model's adversarial training, we compare the translator's performance using two averaging metrics: one reporting the general average from epoch 500 to 4000, while the other informs the maximum 50-epochs moving average among the five sound classes (Fig.12). Besides, we omit the informativity from the 2 epochs immediately after the discriminator update since images generated from these models were frequently fooling the classifiers, which misreported higher informativity, despite the presence of visual artifacts on the images. Thus, the average of the five previous epochs is used instead.

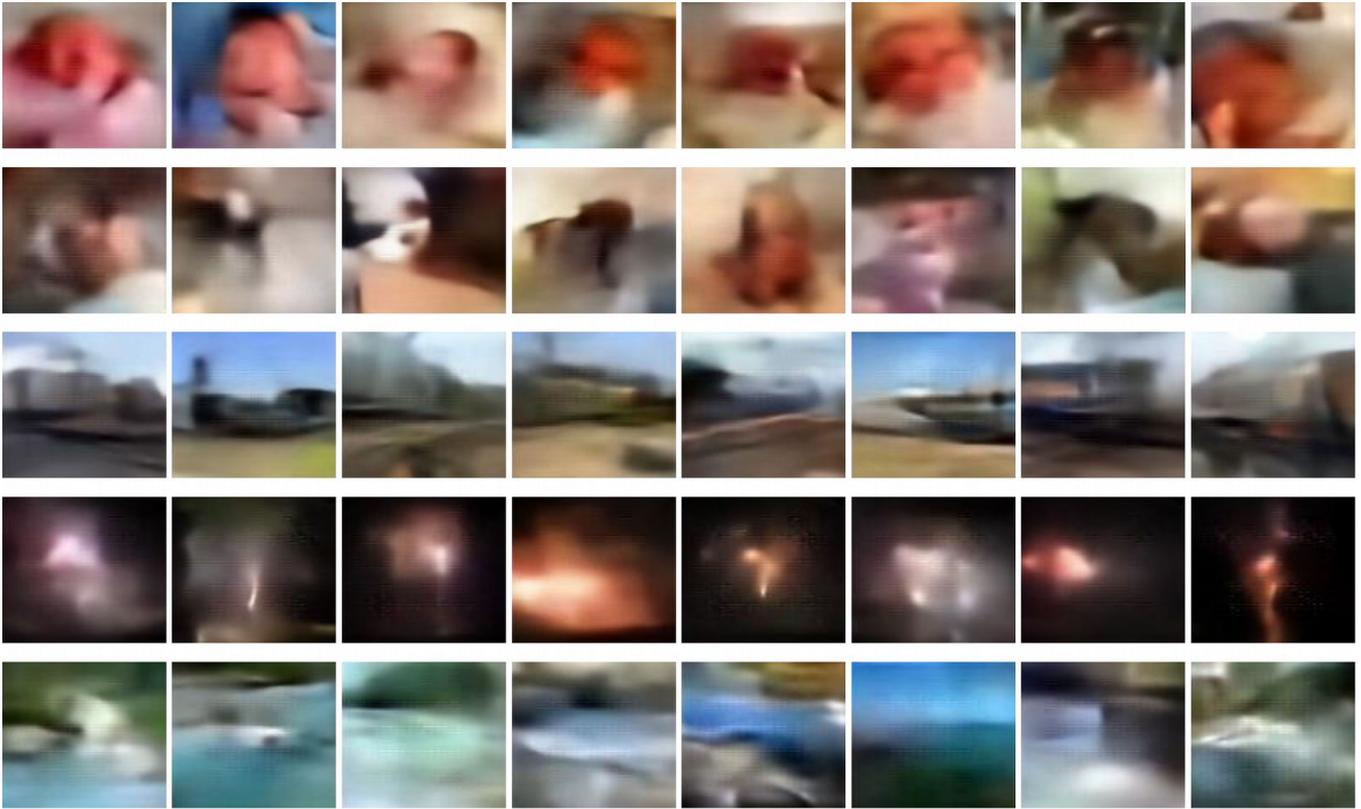

**Fig. 15** Qualitative demonstration of results among the five sound classes, from top to bottom row: Baby cry, Dog, Rail transport, Fireworks, and Water flowing. S2I translation was performed from unknown sounds using a single translator model conditioned on 512-dimension audio embeddings.

The generators training pixel loss is also reported using a 50-epochs moving average (Fig.13). In general, models trained within a broader latent space were able to converge faster and achieve lower pixel losses (Figs. 13 and 14), although this tendency was not verified in extreme dimensions 128 and 2048. In spite of that, informativity decreased when embedding dimension was greater than 512 (Fig.12). Although in a subtle way, this outcome indicates a possible influence of the variation of the audio embedding dimension in the translation performance and the results seem to point out a trade-off between convergence (in the pixel space) and informativity, obtained respectively from higher and lower feature space dimensionality. The higher flow of information across the network seems to have provoked model overfitting, what is difficult to verify since test data losses are uninformative due to the visual decoupling mentioned in Section 5.2. On the other hand, models trained in a more constrained feature space may have poorer convergence to target training data, but, in a general way, they produced more informative images, especially for intermediate dimensions. We hypothesize that this occurs due to the semantic generalization provoked by constraining the information flow between source and target spaces, as demonstrated in previous studies on supervised learning (Alemi et al., 2017), generative adversarial learning (Jeon et al., 2018) (Peng et al., 2020), and domain adaptation (Luo et al., 2019) (Song et al., 2020). In such processes, the reduction of the latent variable dimensionality forces the network to extract essential semantic information. In the case of crossmodal tasks, it helps the translation process on bridging source and target modalities since they are linked by high level semantics. In S2I translation, the concept of an acoustic event is what links aural and visual modalities and all the AudioSet (Gemmeke et al., 2017) structure is based on such concepts. It also worths emphasizing that the proposed S2I translation envisions not only the realism of generated images, but especially their informativity.

Luo et al. (2019) improved the stabilization of GAN training for semantic segmentation through bottleneck constraint. We have managed to overcome major instability issues on long term training using a moving-average discriminator loss, and did not notice any correlation between embedding dimension and training stabilization. Therefore, we assume that the improvement of the performance obtained through the bottleneck constraint is not related to GANs stability. We still cannot state that the bottleneck forced a semantic alignment between modalities and further tests would be needed to confirm the observed tendency, especially by addressing quantification of information along the network. Nonetheless, we highlight the potential of controlling the information flow for solving problems related to semantic alignment involving different modalities and/or domains.

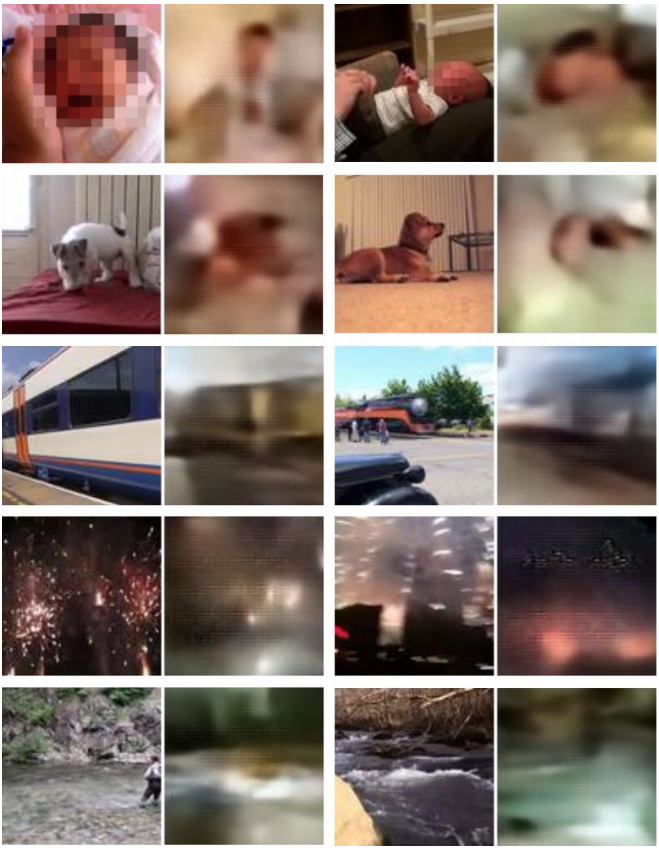

**Fig. 16** Comparison showing a visual decoupling between original and synthetic images among the five sound classes, from top to bottom row: Baby cry, Dog, Rail transport, Fireworks, and Water flowing. S2I translation was performed from unknown sounds using a single translator model conditioned on 512-dimension audio embeddings. The babies' faces in the images on the left were intentionally pixelated to preserve the children's identity.

### 5.4.2. Qualitative Evaluation

All generated images presented in this section are conditioned on sounds from unknown videos and the results demonstrate the translator's achieved generalizability. In Figure 15 we present a selection of translated images from the five sound classes. All images presented in this section, except some shown in Figure 20, were obtained from a single translator model conditioned on 512-dimension audio embeddings. Despite the blurry aspect present on most image areas, which we discuss next, there are identifiable borders and interpretable shapes in all classes and even occasional sharp details. Color coherence is also noticeable, and visual structure in most images is drawn in accordance with real scenes.

Besides, in most images we can see well pictured volumes, with correct light and shadow effects. As mentioned previously, we expected content diversity to be the main challenge we had to address. In fact, fireworks images, which present lower diversity in training data than other classes, showed the best results, not only regarding informativity, but also image sharpness. Although, the abstract visual aspect, inherent in fireworks scenes, may contribute to that impression. On the other hand, translations of dog sounds presented the worst results, which we believe to be due to some characteristics of this class. Since dog sounds are typically short or nearly instant events, it seems that our segmentation of the sound signal was not sufficient to allow a proper modeling of the semantic correspondence between aural and visual modalities for this category.

Apart from dealing with model generalization, we also had to face another important issue regarding the quality of translation. In most generated images, a lack of sharpness was observed, which can be a consequence of using averaged pixel-wise losses, that are known to produce blurry results. Although, these outcomes could also have been provoked by the dropout regularization we used to improve the autoencoder generalization, since it softens the latent feature space. On the other hand, the adversarial loss may have compensated this tendency to a certain extent (Pathak et al., 2016), since blurred images are penalized for having an unreal aspect. Aside from that, S2I translation presents another concern regarding image generation, which is the level of confidence of the translation provided. Despite that we intend to generate images with the best possible quality, we hypothesize that the blurred areas may work as an uncertainty map (Kendall & Gal, 2017) (Sedai et al., 2018). I.e. when the model is not sure about what to draw somewhere in the image, it may produce fuzzy shapes, leaving explicit the inherent uncertainty of the inference. Although we cannot assume that the uncertainty mapping will automatically occur, it is worth investigating the translator's ability to provide such information, either merged with the translated image or in a separated uncertainty layer.

Another characteristic of the translator is that the model successfully produced diverse outputs, as can be seen in the synthetic images shown along this section. However, if compared to the original images that correspond to the audio input, there is still some loss of diversity. Beyond that, in Figure 16, in some examples more than in others, we can notice the aforementioned visual decoupling between the synthetic images and their corresponding ground truth of the test set, which occurred in most translations. On the other hand, we could found exceptions to this rule, and even images translated from unknown sounds may occasionally be similar to the input sound corresponding scenes. In Figure 17, we present a selection of translations that produced such result. Using the currently evaluated model, in most cases presented, the translated images share the original visual structure and also the color, but sometimes the generated image has a mirrored structure compared to the original one (Fig.17 1st row (a) on top and 3rd row (d)). Also, the translated image may match the original visual structure, but colors and textures are different like the mentioned mirrored image (Fig.17 1st row (a)) that pictures a baby crying, and the dog image (4th row (a)), where, despite of having completely different backgrounds, the blurry silhouette on the translated image shows the dog in a quite similar pose when compared to its original pair. Some of the remaining examples present unexpected visual matching in rail transport, fireworks and water scenes, and even a baby crying translation (Fig.17 3rd row (b)) that matches quite precisely both structure and color, except for the baby's clothes. In fact, sound signals carry clues about the surrounding space and scene elements, and this information may be modeled by the translator in order to help picturing the output image. Therefore, what seems to happen by chance, can be revealing learning processes occurring in multimodal intermediate layers of the translator.

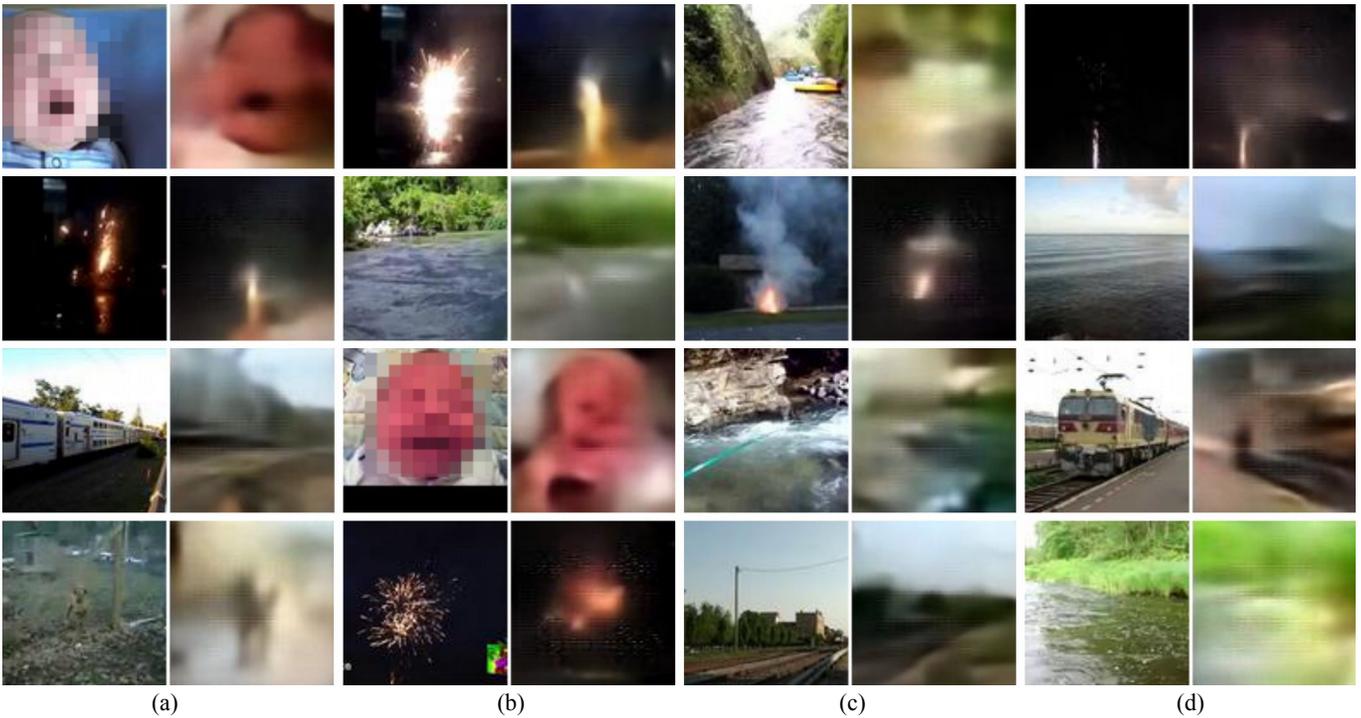

**Fig. 17** Sharing of similar visual structure (and also color in some cases) between synthesized images and input sound corresponding images. S2I translation was performed from unknown sounds using a single translator model conditioned on 512-dimension audio embeddings. The babies' faces in the original images were intentionally pixelated to preserve the children's identity.

Regarding badly translated images, such results occur due to different reasons. For instance, some acoustic events may not have been well modeled, which will result in strange or almost abstract images, as occurred with most dog images.

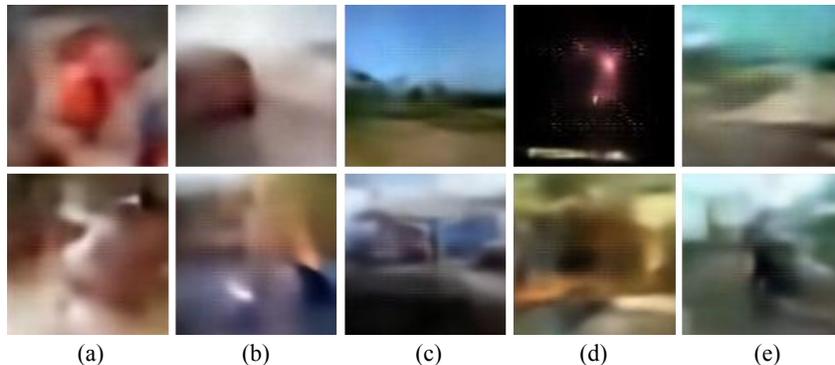

**Fig. 18** Typical examples of badly translated images from the five sound classes, from left to right column: (a) Baby cry, (b) Dog, (c) Rail transport, (d) Fireworks, and (e) Water flowing. S2I translation was performed from unknown sounds using a single translator model conditioned on 512-dimension audio embeddings.

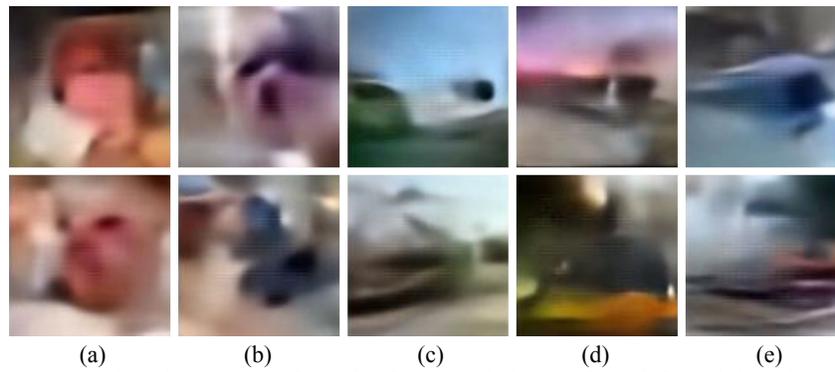

**Fig. 19** 'Creative' outputs from the five sound classes, from left to right column: (a) Baby cry, (b) Dog, (c) Rail transport, (d) Fireworks, and (e) Water flowing. S2I translation was performed from unknown sounds using a single translator model conditioned on 512-dimension audio embeddings.

Other times the model just mistranslated the input sound and, for example, generated a water fall scene for a passing train sound, or vice versa. Also, problems related to the dataset may interfere in translation quality. For instance, regarding fireworks sounds, frames from the original videos of the training data frequently contained white subtitles in the lower area, which ended up been modeled as part of the sound emitting source, and sometimes appears on generated images in the form of horizontal luminous lines over the dark background (Fig.18 1st row (d)). Apart from the informative/non-informative images perspective, we identified the following types of translated images:

- Defective – Images that may even be informative, but that contain elements wrongly pictured, regarding color, luminosity, size and/or position.
- Incomplete – Images that may be informative to some extent, but that miss some essential part of the element of interest. Or, despite having a coherent surrounding scene, the sound emitting source is omitted.
- Artifactual – Non-informative images that are rather abstract, consisting basically of unrecognizable forms.
- Implausible – Images that can sometimes be informative, but that contain awkward elements.
- Surreal – Images that may be informative to some extent, but that look curious or rather fantastic.
- Creepy – Images that may be partially informative, but that picture parts of living beings in a harrowing way, or that contain ghostly or alien-looking elements. These images could sometimes also be considered defective or surreal, depending on the case.
- Multi-informative – These images picture elements from two or more sound classes.

With respect to defective, incomplete and artifactual images, such outputs occurred with different frequencies, and sometimes can significantly compromise the informativity of the translation. For instance, a typical defective image occurs when parts of the image are just wrongly pictured like in Figure 18 (1st row (a)), where the baby's eyes region are brighter than the rest of the face, when usually the opposite is likely to happen. However, what is clearly a bad output, can also indicate the capacity of the translator to separately model image semantic parts. Another example of a defective image can be seen in Figure 18 (1st row (e)), where the water scene visual structure appears to be upside down. And what could be the corner of a rocky beach with clear green water ends up looking like an abstract image. Respect to incomplete images, in the case of rail transport sounds for instance, the translator may output an interpretable picture of a landscape, which is coherent with the acoustic event in question, but we cannot see any train, nor even the rail way (Fig.18 1st row (c)). Another example of incomplete image can be seen in Figure 18 (1st row (b)), a potentially informative picture of a dog that gets compromised due to not having any element that indicates where is the head of the animal, and the image ended up looking rather abstract. Considering the entire dataset, abstract images occurred more frequently in the translation of dog sounds than any other class, but nonsense pictures may show up anytime. The most common outputs of this type are artifactual images, of which we show some examples in Figure 18 (2nd row (a to e)).

As mentioned in Section 3, since the translator must generate images based only on the input sound without any visual information of the original scene, we realized that the crossmodal generation performed in S2I translation implies addressing the problem of computational imagination. Despite that our goal is to produce realistic images, there is an inherent 'creativity' necessary to enable the translator to 'imagine' a complete scene. The fact is that we have found that sometimes the translator was able to produce rather 'creative' results. Although these outcomes were not intended, they were expected to happen, and they are an indicator of some level of 'creativity' achieved, since they demonstrate that the translator was capable of using learned visual features to generate original forms. Surreal, implausi-ble and creepy images can be examples of 'creative' outputs, as shown in Figure 19. Sometimes the 'creativity' of the model will compromise the informativity of the image, and other times it just adds a bit of fantasy to the picture. With respect to implausible images, despite that such images may sometimes be informative, the presence of awkward elements ends up diverting attention from relevant information. In Figure 19 (1st row (a)) we show an example of this, that is an implausible picture of what seems to be a baby reading a pink-covered book. Also from baby crying sounds, we can see in the image below (2nd row (a)), a bit creepy picture of what looks like an alien's face. Regarding surreal images, also in Figure 19 (1st row (b)), we can see what was supposed to be a dog, since the image is translated from dog sounds, but, maybe due to the shape of the head, it looks more like a fantastic creature with a furry face. The image below (2nd row (b)), also translated from dog sounds, shows a figure that is a sort of dark gray dummy sitting on the floor of some indoor environment. From train sounds, we also obtained some 'creative' outputs. On the same figure (1st row (c)), we find a kind of white spaceship landed over a grass field with blue sky on top, but the picture could be also of a white face with black eyes. The image below (2nd row (c)) is also curious, showing what seems to be an outdoor scene in daylight, with some sort of machinery over a gray pavement. From fireworks sounds, we obtained some colored compositions, one that looks more like a sunset scene (1st row (d)) and another that is rather abstract with something on the top that could be a fireworks flash (2nd row (d)). Finally, respect to water sounds, the translator pictured a sort of water-made creature, or maybe a blue fish (1st row (e)). While the image below (2nd row (e)) appears to show a face surrealistically merged with the landscape.

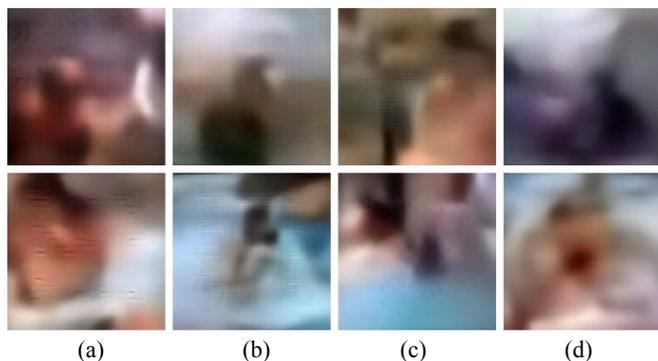

(a) (b) (c) (d)

**Fig. 20** Multi-informative images translated from sounds of people and fireworks (1st row on top), and from sounds of people and water (2nd row). S2I translation was performed from unknown sounds using seven different translator models conditioned on the following audio embedding dimensions: 128 (1st row (a and b)), 512 (1st row (c and d), and 2nd row (a, b and c)), 1024 (2nd row (d)).

Furthermore, the translator was capable of generating multi-informative outputs, synthesizing more than one acoustic event in a single image. However, this kind of inference rarely occurred. As far as we could verify through subjective evaluation using 22 different models, the translator generates, in average, two multi-informative images from the entire test set of 6825 sounds. The easiest to spot are the ones that picture people inside the scene. In Figure 20, we show multi-informative images translated from sounds of people and fireworks (1st row on top) and from sounds of people and water (2nd row). The images presented were obtained from S2I translation performed from unknown sounds using seven different translator models. The occurrence of such images is so rare probably due to the fact that the dataset was not prepared for multiple sounds inference. As explained before, the data used for training the translator was pre-cleaned only to ensure that the sound related element/event of interest was present both in aural and visual modalities. But since the audiovisual content is diverse, the video segments may contain other acoustic events occurring simultaneously. The problem is that there is no guarantee that these extra acoustic events are also present in both modalities of the training data. To prevent any incoherence of this kind, it would be necessary to ensure that all acoustic events of the audio stream had their visual elements represented in the corresponding video frames, which could filter out the data excessively, making it impossible to train the translator. Nonetheless, the obtained multi-informative images are an indicator that the unsupervised approach allowed the translator to share features freely and spontaneously generated images spotting the presence of people in the acoustic scenes. The voice is the most common audible indicator of human presence, whether through speech (Fig.20 1st row (a, c, d), 2nd row (a, b, d)), shout (1st row (a), 2nd row (a)) or singing (1st row (b)). But in the case of the image in 2nd row (c), also in Figure 20, human presence was detected by a loud and fussy dog paddle swim performed by a man in the river. In this case, one single sound carries the information about two events, the man swimming and the water movement.

## 6. Conclusion and Future Work

We conducted an exploratory study designing, training and testing an end-to-end S2I translator with a deep dense generator architecture, presenting details of the model, the heuristics of the approach and an evaluation of the results of a S2I translation experiment. We also presented a solution employing informativity classifiers to report the translator's performance. As far as we know, this is the first work to address unsupervised S2I translation with such a level of audiovisual content diversity. Despite that the translator often produced non-informative outputs, our model was capable of generating, in average, more than 14% of interpretable and semantic coherent images. Among these, some even presented a visual structure similar to the input sound corresponding images. In addition to the achieved informativity, the translator was able to produce visually diverse results. We also have found that the translator sometimes produced 'creative' results picturing original forms, and, less frequently, spontaneously generated multi-informative images. Furthermore, we compared the performance of five different S2I translator models regarding variation of the audio embedding space dimensionality. In a subtle way, the results indicate a trade-off between pixel space convergence and informativity, obtained respectively from higher and lower feature space dimension. We believe that the increased informativity of generated images is due to semantic generalization provoked by constraining the information flow between source and target spaces. Although, further studies addressing quantification of information along the network would be needed to confirm our assumption. Besides, other explanations for the influence of bottleneck variation over performance must be considered. Apart from the control of information flow, the use of a deeper and dense architecture was decisive to the improvement of the translator's generalization. These and other solutions that we have presented allowed us to overcome the problem to a certain extent. We highlight the necessity to further explore the characteristics of the networks that can interfere in the generalization of GANs applied to crossmodal tasks. We also encourage approaches to optimize the model using perceptual losses jointly with the adversarial loss and the averaged pixel-wise loss. Moreover, finding feasible solutions to address a broader sonic universe is a key step for taking forward the research on S2I translation.

## Acknowledgements

The authors are thankful to Santiago Pascual for his advice on the implementation of GANs. We also thank Josep Pujal for his support in using the computational resources of the Signal Theory and Communications Department at the Polytechnic University of Catalonia (UPC).